\documentclass[fleqn,10pt]{wlscirep}
\usepackage[utf8]{inputenc}
\usepackage[T1]{fontenc}
\usepackage{braket}
\title{Addressing Quantum's ``Fine Print'': \\
\large State Preparation and Information Extraction for\\ Quantum Algorithms and Geologic Fracture Networks}

\author[1,*,+]{Jessie M. Henderson}
\author[2,+]{John Kath}
\author[1]{John K. Golden}
\author[2]{Allon G. Percus}
\author[3]{Daniel O'Malley}
\affil[1]{Los Alamos National Laboratory, CCS-3, Los Alamos, NM 87545, USA}
\affil[2]{Claremont Graduate University, Institute of Mathematical Sciences, Claremont, CA 91711, USA}
\affil[3]{Los Alamos National Laboratory, EES-16, Los Alamos, NM 87545, USA}
\affil[*]{jessieh@lanl.gov}
\affil[+]{These authors contributed equally to this work.}


\begin{abstract}
    Quantum algorithms provide an exponential speedup for solving certain classes of linear systems, including those that model geologic fracture flow.
    However, this revolutionary gain in efficiency does not come without difficulty.
    Quantum algorithms require that problems satisfy not only algorithm-specific constraints, but also application-specific ones.
    Otherwise, the quantum advantage carefully attained through algorithmic ingenuity can be entirely negated.
    Previous work addressing quantum algorithms for geologic fracture flow has illustrated core algorithmic approaches while incrementally removing assumptions.
    This work addresses two further requirements for solving geologic fracture flow systems with quantum algorithms: efficient system state preparation and efficient information extraction.
    Our approach to addressing each is consistent with an overall exponential speed-up.
\end{abstract}
\begin{document}

\flushbottom
\maketitle

\thispagestyle{empty}

\section*{Introduction}

Quantum algorithms promise to revolutionize the solving of linear systems, which are essential components of problems in medicine, finance, urban development, and nearly any field that can be classified as a natural or applied science [\citenum{boyd2018introduction}].
Several quantum algorithms  [\citenum{harrow2009quantum,ambainis2012variable,childs2017quantum,wossnig2018quantum,subasi2019quantum,costa2021optimal,omalley22woodbury}] can provide a provable exponential speedup over classical linear solvers. However, remarkable though such gains are, they do not come without cost, nor without complication [\citenum{aaronson2015read}].
Problems of interest must be curated to satisfy algorithm-specified constraints. Moreover, the quantum algorithms themselves must account for the complexity of transporting information to and from the quantum computer via processes that bear little resemblance to classical counterparts [\citenum{aaronson2015read}].
Otherwise, all theoretical intrigue aside, quantum linear-systems algorithms become toothless: we know the algorithm could compute the solution exponentially faster than possible classically, but we can neither supply the problem nor extract the solution efficiently enough to benefit from this speedup [\citenum{nielsen2010quantum}].
These difficulties are further complicated by context-specific considerations; for example, it would be pointless to prepare and solve a
linear system if we cannot extract information that is of practical relevance for the problem at hand.

Consequently, efficient approaches for both preparing the system to be solved and extracting useful information from quantum computers are as important as quantum algorithms themselves.
This paper addresses these issues within the context of quantum algorithms for geologic fracture networks.\footnote{The approaches are fairly general and thus likely have utility for applications beyond geologic linear systems. Nonetheless, this work explicitly addresses only the realm of geologic fracture problems.}
Linear systems representing fracture networks are too large to solve in their entirety with even the most sophisticated classical approaches [\citenum{hyman2015dfnworks,mills2007simulating}], and reducing problem size requires methods such as upscaling, which supply only approximate solutions that may neglect important features of the network.
For example, when small fractures are neglected, a network exhibiting percolation---complete connectivity of a fracture region---might no longer manifest that effect [\citenum{omalley2016water}].
Such modelling issues make geologic fracture problems a prime candidate for benefiting from the speedup provided by quantum algorithms, so long as we can satisfy the algorithmic constraints and provide efficient state preparation and information extraction.
Previous work has addressed solving fracture flow problems with quantum algorithms  [\citenum{omalley2018approach,henderson2021interrogating,sahimi2022potential,henderson2023quantum,greer2023early}] while making assumptions about the auxiliary issues.
Ref. [\citenum{golden2022preconditioning}] then addressed the requirement of well-conditioned matrices by developing effective preconditioning permitting quadratic speedup for systems representing geologic fracture problems.
Here, we address two further constraints of efficient state preparation and solution extraction.

The remainder of the paper proceeds as follows. We first provide an introduction to modelling geologic fracture networks with linear systems, including the relationship to quantum algorithms.
We then present methods for state preparation and information extraction that have acceptable complexities and that are readily usable by human developers, including on noisy near-term quantum hardware.\footnote{Although the Harrow-Hassidim-Lloyd (HHL) algorithm that we discuss below is generally not appropriate for near-term hardware, this suggests that our approaches for state preparation and information extraction could perhaps be applied to other quantum linear solvers more apt for today's NISQ-era machines.}
Finally, we conclude with a brief discussion of future work.

\section*{Background}\label{sec:background}
\subsection*{Quantum Algorithms for Geologic Fracture Networks}
Simulating geologic fracture networks is one of the most challenging problems in geophysics, in part because of the large range over which fractures exist [\citenum{fountain2005fractures,davies1999role,viswanathan2022fluid,laubach2019role}].
Systems modelling fractures with sizes between $10^{-6}$ and $10^4$ meters cannot be solved accurately in their entirety on classical machines, and they sometimes cannot be accurately upscaled either.
Specifically, information lost during upscaling pertains to small fractures that can have a critical effect on the fracture network; for example, the smallest fractures can determine whether the network crosses a percolation threshold, which has a substantial impact on fluid flow [\citenum{omalley2016water}].

Quantum algorithms for solving linear systems are not burdened by the same constraints as their classical counterparts. The properties of quantum mechanics endow them with a fundamentally different physics---and thus a fundamentally different mathematics---that allows for efficiently solving problems that cannot be solved on classical computers using a reasonable amount of memory or time [\citenum{nielsen2010quantum}].
Previous work has both explained and demonstrated use of quantum algorithms for solving linear systems problems in the geologic fracture realm  [\citenum{omalley2018approach,henderson2021interrogating,sahimi2022potential,henderson2023quantum}], but with the caveat that future work would need to explore efficient mechanisms of introducing the problem to the computer and extracting meaningful information from the solution.
In this work, we consider those issues for pressure-identification problems of the form $\Delta \cdot (k\Delta h) = f$, where $k$ is permeability, $f$ is a fluid source or sink, and $h$ is the pressure to be computed.
These problems can be discretized and written as $A\mathbf{x}=\mathbf{b}$, where the pressure for each discretized node is stored in $\mathbf{x}$.
Then, quantum algorithms for solving linear systems can compute a normalized vector that is proportional to that solution  [\citenum{harrow2009quantum}].

One such algorithm is the Harrow-Hassidim-Lloyd (HHL) algorithm, which was the first for solving linear systems with quantum circuits.
It provides an exponential speedup over classical algorithms under certain conditions, including constraints on matrix sparseness and condition number [\citenum{harrow2009quantum}].
Because geologic fracture flow systems can be made to satisfy such conditioning requirements  [\citenum{golden2022preconditioning}], and because they are unavoidably large in their complete form, such systems are ideal candidates for the algorithm, assuming that we can efficiently specify the problem and extract the solution.
Approaches to these information-transfer tasks are not algorithm-independent; the details can depend upon how a given quantum algorithm prepares the solution vector $\mathbf{x}$.
However, there are often similarities in quantum linear systems algorithm structure that would make state-preparation and information-extraction approaches inter-algorithmically applicable.
In this work, we will directly consider only the HHL algorithm while acknowledging that our methods for state preparation and information extraction may be more broadly applicable.

\subsection*{Brief Introduction to HHL}
The HHL algorithm prepares a solution proportional to that of the $N\times N$ system $A\mathbf{x} = \mathbf{b}$  [\citenum{harrow2009quantum}].
A single execution of the algorithm has a complexity of $O(\log{(N)}s^2\kappa^2/\epsilon)$, where $N$ is the size of system, $s$ is the sparseness of the matrix, $\kappa$ is the condition number of the matrix, and $\epsilon$ is the additive error within which the system is solved  [\citenum{harrow2009quantum,chakrabarti18}].
As with most quantum algorithms, HHL is at once fairly simple in qualitative terms and quite subtle in quantitative ones.
While this work is not intended as a detailed introduction to HHL, it is worth briefly describing the overall algorithm and highlighting a few relevant points.\footnote{Given the algorithm's wide applicability and dramatic speedup, several works look to provide detailed---yet accessible---treatments of the algorithm; for more information, please see Refs. [\citenum{morrell2021step}] or [\citenum{duan2020survey}].}

\begin{figure}[ht]
\centering
\includegraphics[width=0.7\linewidth]{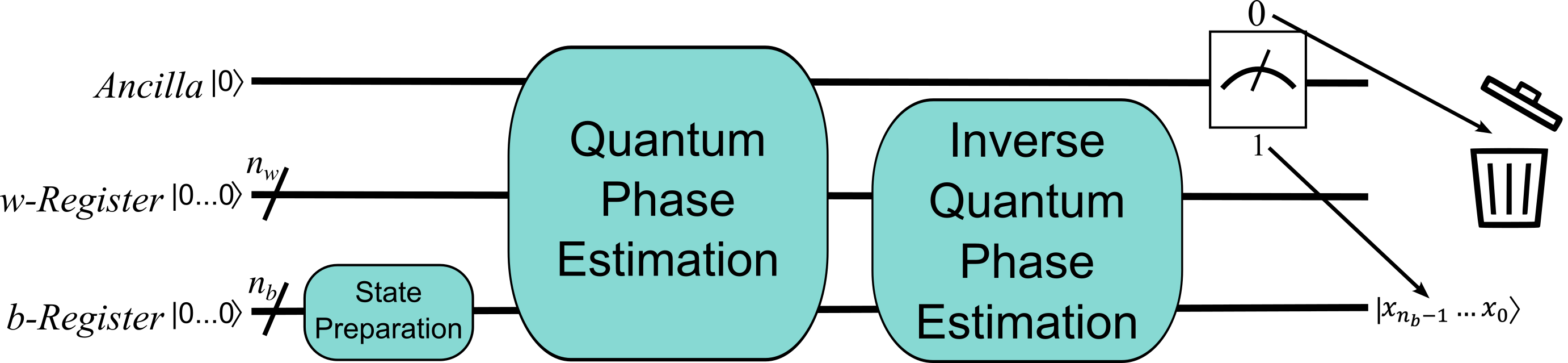}
\caption{A block diagram of HHL. After a separate, \textit{non-HHL} algorithm has prepared the right-hand-side vector, $\mathbf{b}$, HHL applies QPE to extract the eigenvalues and eigenvectors from the matrix, $A$. Then, the algorithm prepares the eigenvalues for extraction from their entangled state using inverse QPE before measuring an ancilla qubit to determine if the operation was successful. A measurement of $1$ indicates that $\ket{x} \sim \mathbf{x}$ is stored in the $b$-register, while a value of $0$ indicates a probabilistic circuit failure, the likelihood of which is included in the complexity of HHL via $\epsilon$.}
\label{fig:HHL}
\end{figure}

HHL prepares a normalized solution to $\mathbf{x}$ by leveraging the fact that $\mathbf{x} = \sum_{i=1}^N \lambda_i^{-1}b_i\ket{u_i}$, where $\ket{u_i}$ is the $i$th eigenvector of $A$ and $\lambda_i$ is its associated eigenvalue.
Specifically, as shown in the block diagram of Figure \ref{fig:HHL}, the algorithm requires two registers of qubits and an ancilla qubit.
The $b$-register begins as storage for normalized values proportional to those in the right-hand side vector of $\mathbf{b}$, and if the ancilla is measured as $1$, then it ends storing $\ket{x}$, which is a normalized vector proportional to $\mathbf{x}$.
HHL therefore requires an efficient mechanism for converting the $b$-register (with $n_b$ qubits) from the fiduciary state of $\ket{0...0}$ to a state in which each qubit holds two values of the normalized $\mathbf{b}$: $\ket{bReg_0} = b_0\ket{0} + b_1\ket{1}, \ket{bReg_1} = b_2\ket{0} + b_3\ket{1}, ..., \ket{bReg_{n_b}} = b_{2^{n_b}-2}\ket{0} + b_{2^{n_b}-1}\ket{1}$.

The $w$-register is a working register that is used to store intermediate values throughout the computation.
Specifically, it is used during the subroutines of Quantum Phase Estimation (QPE) and Inverse QPE, which identify and isolate the eigenvalues $\lambda_i$ of $A$ alongside a normalization constant that makes the final state proportional to (rather than equal to) $\mathbf{x}$.
Consequently, the information about $A$ necessary to solve the linear system is encoded in the QPE subroutine, and HHL's efficiency requires that $A$ be well-conditioned and sparse for this information-transfer process to avoid a complexity greater than that of the entire solve.

Finally, the ancilla qubit determines whether $\ket{x}$ is properly stored at the end of the algorithm; measuring it decouples the solution to the system ($\ket{x} = \gamma\sum_{i=1}^N \lambda_i^{-1}b_i\ket{u_i}$, where $\gamma$ is a normalization constant) from associated `garbage' information ($\ket{x} = \sum_{i=1}^N \sqrt{1 - \frac{\gamma^2}{\lambda_i^2}}b_i\ket{u_i}$) that is added throughout the computation.
Thus, HHL requires multiple circuit executions to account for when the ancilla is measured as a zero [\citenum{chakrabarti18}].

\begin{figure}[ht]
\centering
\includegraphics[width=0.5\linewidth]{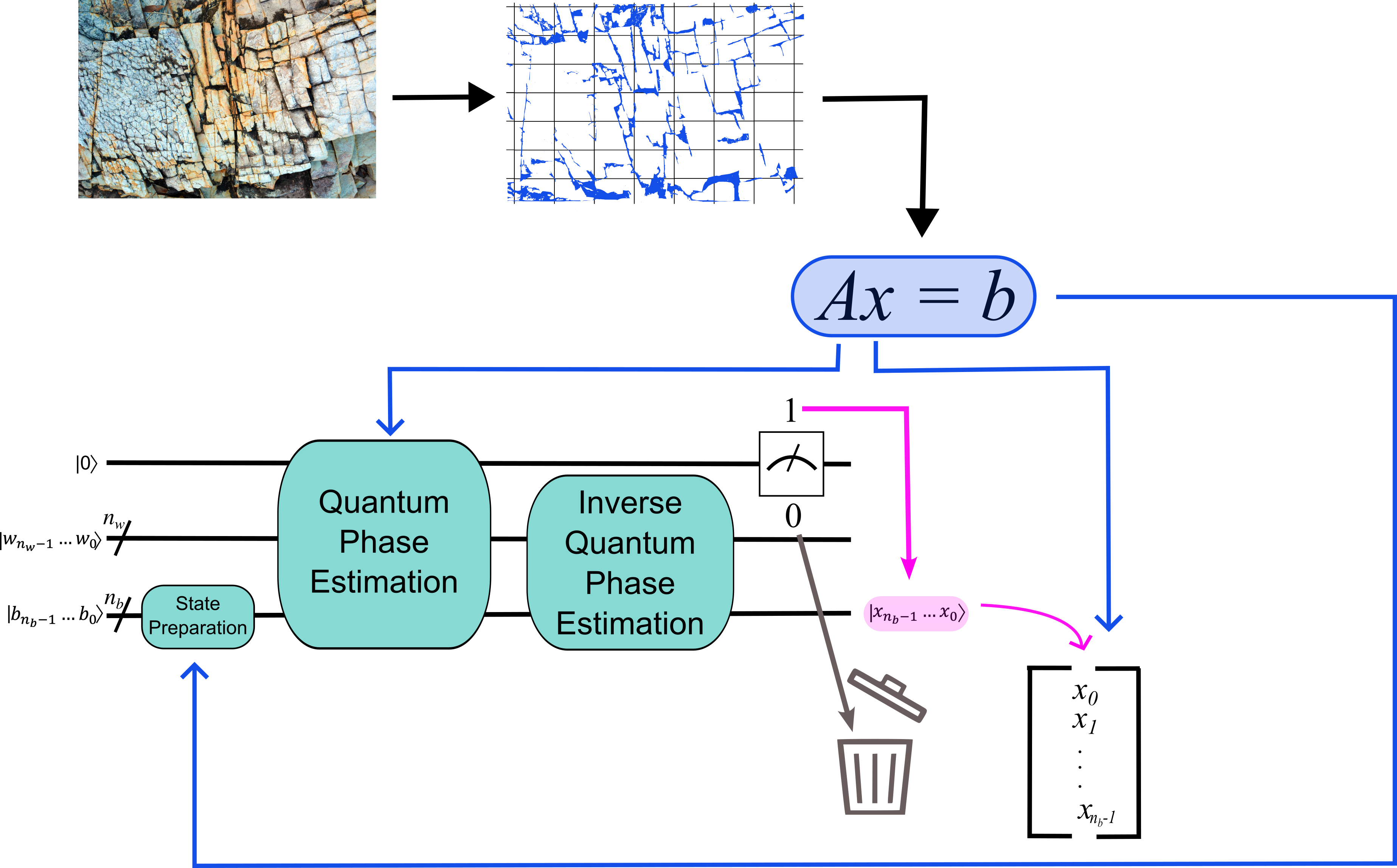}
\caption{A schematic illustrating how information is transferred from a geologic fracture flow problem into an HHL circuit that solves for the pressure at each discretized node.}
\label{fig:HHL_for_geosciences}
\end{figure}

Figure \ref{fig:HHL_for_geosciences} illustrates the relation between a geologic fracture flow problem and a circuit implementing HHL.
First, the region of interest for a particular problem (say, the $\Delta \cdot (k\Delta h) = f$ of above) is discretized to form $A$ and $\mathbf{b}$.
Then $A$ determines the parameters in the fixed structure of a QPE subroutine.
A state preparation procedure encodes the values of $\mathbf{b}$ in the $b$-register, and both the ancilla qubit and the qubits in the $w$-register begin as $\ket{0}$.
If, at algorithm completion, the ancilla is measured as $1$, then the problem solution---representing the pressure at each node of the discretization---is stored in what began as the $b$-register: $\ket{bReg_0} = x_0\ket{0} + x_1\ket{1}, \ket{bReg_1} = x_2\ket{0} + x_3\ket{1}, ..., \ket{bReg_{n_b}} = x_{2^{n_b}-2}\ket{0} + x_{2^{n_b}-1}\ket{1}$.

Useful though such conceptual schematics are, they provide no guidance on efficiently encoding information into the $b$-register or extracting the eventual solution.
As Scott Aaronson recognizes in his oft-cited remarks on the ``fine print'' of quantum algorithms [\citenum{aaronson2015read}], this is in part because such determinations are often very difficult problems in and of themselves, and indeed can be so difficult that the issue of transferring information to and from the quantum computer bars use of quantum approaches that would otherwise be preferable to classical variants.
For example, obtaining the entirety of $\ket{x}$ requires $O(N)$ measurements, which requires running the entire HHL circuit $O(N)$ times, thus negating the complexity benefit provided by HHL in the first place.
The next subsection introduces two subroutines that we will use to address this issue.

\subsection*{The Swap and Hadamard Tests}
\begin{figure}[t]
\centering
\includegraphics[width=0.7\linewidth]{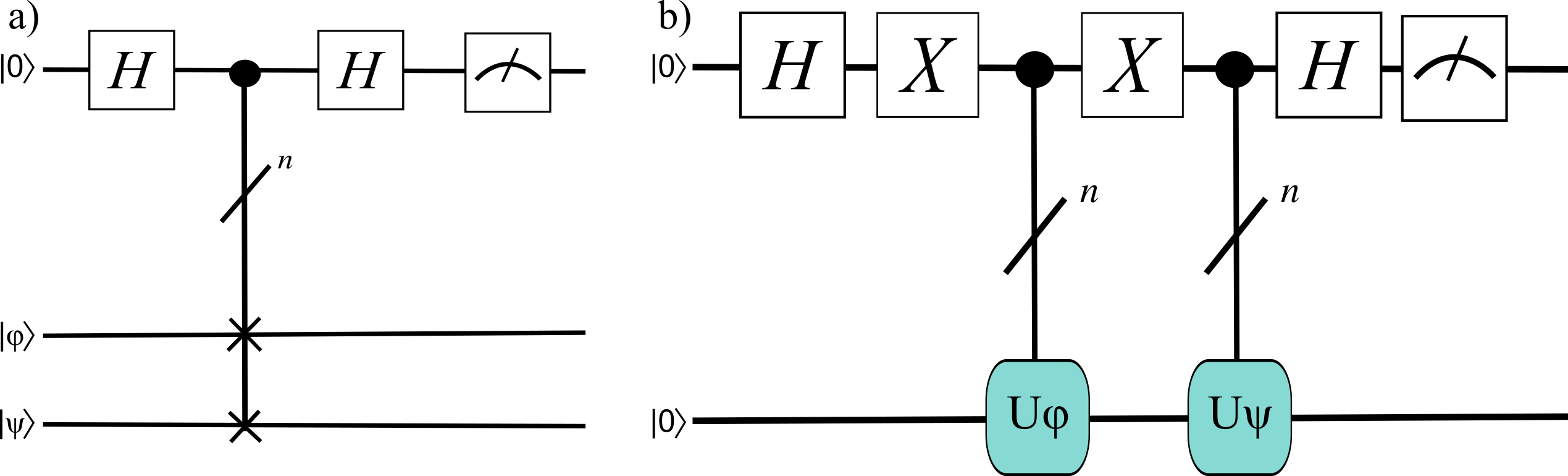}
\caption{Subfigure A illustrates the swap test for two $n$-qubit registers, $\ket{\phi}$ and $\ket{\psi}$, while Subfigure B illustrates the Hadamard test for two $n$-qubit states that can be efficiently prepared via the gates $U_\phi$ and $U_\psi$.}
\label{fig:swap_and_hadamard_test}
\end{figure}

The swap test is a straightforward quantum subroutine that obtains the overlap between two $n$-qubit quantum states using $n$ controlled-swap gates, two Hadamard gates, and one measurement  [\citenum{barenco1997stabilization,buhrman2001quantum}].
The circuit must be run $O(1/\epsilon^2)$ times to obtain a solution that is within a specified additive error, $\epsilon$, of the true overlap.
Subfigure A of Figure \ref{fig:swap_and_hadamard_test} illustrates the circuit structure; by measuring the single qubit and obtaining the probability that it is $0$, the inner product between the two registers is given as $|\braket{\phi|\psi}|^2 = 1 - 2p(0)$.
It is thus worth noting that the swap test is capable only of determining the magnitude of the inner product and not its sign; the slightly more complicated Hadamard test addresses this, as illustrated in Subfigure B of Figure \ref{fig:swap_and_hadamard_test}.
Specifically, the Hadamard test for computing inner products uses two Hadamard gates, two Pauli-$\mathbf{X}$ gates, and two controlled-$U$ gates for unitaries $U$ that prepare the states whose overlap is to be computed  [\citenum{cleve1998quantum}].
For two $n$-qubit registers, the complexity is still $O(1/\epsilon^2)$, and the inner product is given by $\text{Re}[\braket{\phi|\psi}] = 2p(0) - 1$.\footnote{The imaginary component can be computed with a slight adjustment, but since geologic fracture network problems do not require complex values, we will not consider that here.}
Below, we will utilize these subroutines to efficiently extract the average pressure from a user-specified set of nodes after solving for the pressure in all nodes.
But first, we describe efficiently preparing the $b$-register.

\section*{Efficient State Preparation}\label{sec:efficient_state_preparation}

One of HHL's critical assumptions is availability of an efficient method for preparing the quantum state of the $b$-register.
The preparation of a general state $\mathbf{b}$ can require a computational effort of $\Omega\left(2^{n_b}\right)$, negating the advantage of a quantum computation.
Our approach, based on the algorithm of Gleinig and Hoefler [\citenum{niels_torsten_aeafsqsp}], enables efficient generation of the quantum state $\mathbf{b}$ specifically tailored to subsurface flow problems.

For our fracture network problems, $\mathbf{b}$ encodes the boundary conditions.
We will consider examples with a combination of Dirichlet boundary conditions, which specify pressure, and Neumann boundary conditions, which specify flux.
Specifically, we present two scenarios that arguably represent the two most common scenarios considered when modelling subsurface flow.

\subsection*{Pressure Gradient}

\begin{figure}[ht]
\centering
\includegraphics[height=0.3\linewidth]{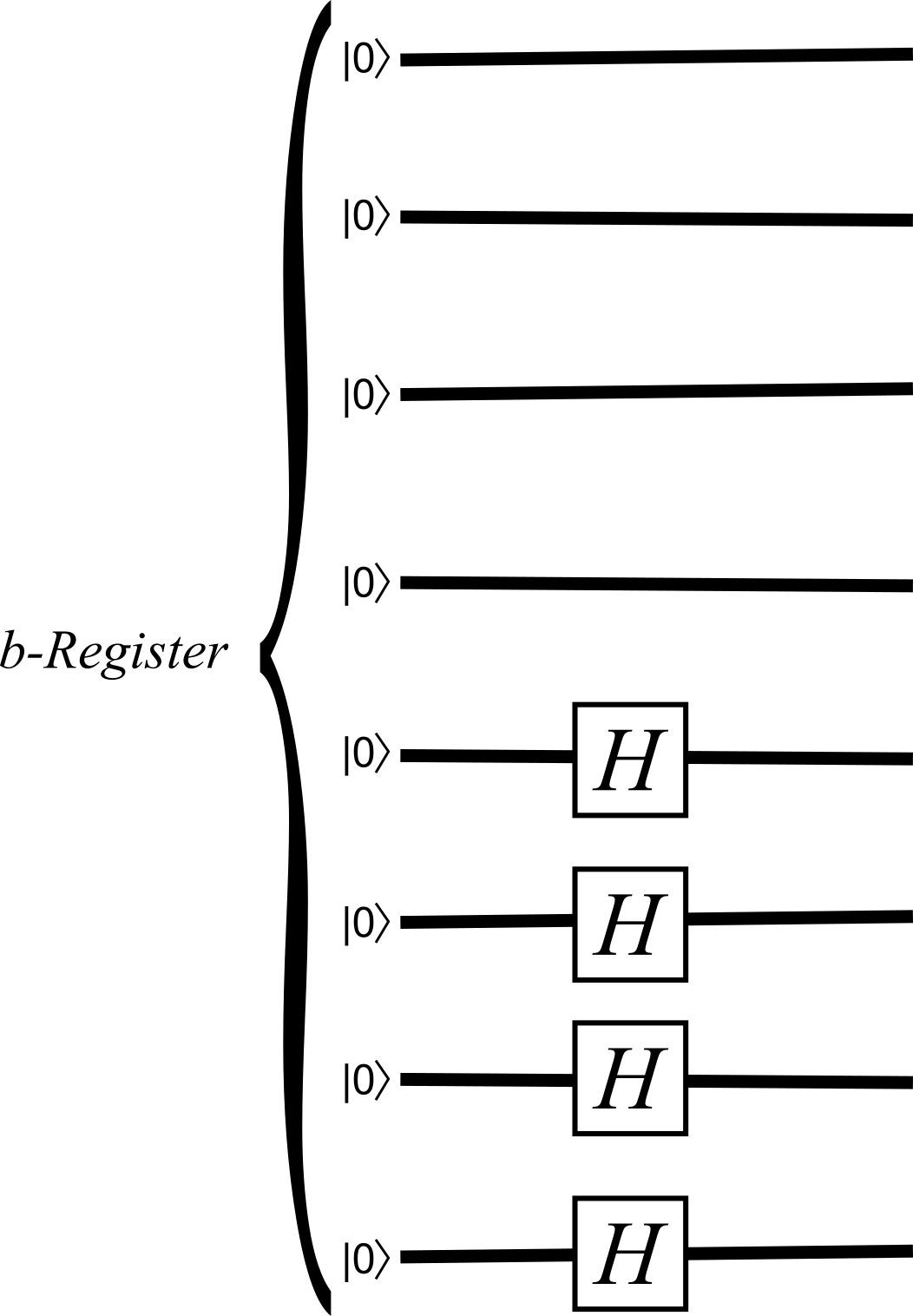}
\caption{A quantum circuit that prepares the state of $\mathbf{b}$ for a fracture flow problem with a pressure gradient where ${n_b}=8$. Note that $\mathbf{b}$ has $2^{n_b/2}$ nonzero entries.}
\label{fig:dirichlet_bc_ex}
\end{figure}

In the first scenario, we consider Dirichlet boundary conditions where the left boundary experiences high pressure while the right boundary has low pressure, and we impose Neumann boundary conditions with zero flux at the top and bottom.
In this case, preparing the state of $\mathbf{b}$ is straightforward.
Without loss of generality, assume that ${n_b}$ is even, and let $m={n_b}/2$.
The $b$-register starts in a state of $2 m$ zero qubits ($\left|0^m\right\rangle\left|0^m\right\rangle$), and we apply Hadamard gates to the second $m$ qubits to put them in a uniform superposition, giving
$$
|0^m\rangle\frac{1}{\sqrt{2^m}} \sum_{i \in\{0,1\}^m}|i\rangle .
$$
This produces a circuit with ${n_b}/2$ gates.
Figure \ref{fig:dirichlet_bc_ex} depicts preparing $\mathbf{b}$ under a pressure gradient when ${n_b}=8$.

\subsection*{Fluid Injection}

The remainder of this section discusses the second scenario.
We consider Dirichlet boundary conditions where the left and right boundaries maintain zero pressure.
We again impose Neumann boundary conditions with zero flux at the top and bottom.
Additionally, we introduce the concept of injecting or extracting fluid at a small number of sites---\textit{i.e.}, wells---in the middle of the domain, which is comparable to Neumann boundary conditions that specify a flux.
It is worth emphasizing that, in practice, the number of wells considered in a given simulation is constant and typically small, meaning that the number of wells does not increase as the simulation mesh is refined.

Because the number of wells is small, the vector $\mathbf{b}$ is sparse, meaning it has few nonzero entries, with the total number of non-zero entries equal to the number of injection/extraction wells.
In general, a circuit for preparing an $n_b$-qubit quantum state is a sequence of $O(2^{n_b})$ gates, where multicontrolled operations are used for readability.
The method of Ref. [\citenum{niels_torsten_aeafsqsp}] exploits the sparsity of $\mathbf{b}$,
taking a classical specification of a quantum state $\phi$ with $W$ nonzero coefficients as input and producing, in polynomial time, a polynomial-size circuit $C$ that maps the initial state $\left|0^{n_b}\right\rangle$ to $\phi$.
Note that $W$ is both the number of nonzero coefficients and the number of injection/extraction wells.

The algorithm works by enabling efficient transformation of quantum states.
These transformations range from basis states to zero states using NOT gates to more complex scenarios involving superpositions of basis states.
The key challenge is preventing the ``splitting'' of basis states during transformation.
Ref. [\citenum{niels_torsten_aeafsqsp}] relies on two crucial techniques:
\begin{enumerate}
    \item Controlling the merging of basis states with no more than $O(\log W)$ control bits.
    \item Using efficiently-implementable gate sequences for multicontrolled operations.
\end{enumerate}

The circuit is generated from a gate library that includes controlled-NOT (CNOT) gates and single-qubit T gates.
Using this method, the state can be prepared using only $O(W {n_b})$ CNOT gates and $O(W \log W +{n_b})$ single qubit gates.
Since $\log W$ is itself $O(n_b)$, this results in a total number of gates that is $O(W n_b)$, rather than of exponential complexity.

\subsubsection*{Application to Fracture Networks}
The polynomial-sized quantum circuit produced for preparing a sparse state illustrates this algorithm's practical relevance for solving geologic fracture flow problems with quantum algorithms.
The following examples demonstrate efficient state preparation for fracture networks.
First, we consider a simple case consisting of two intersecting fractures and two injection/extraction sites.

\begin{figure}[ht]
\centering
\includegraphics[width=0.8\linewidth]{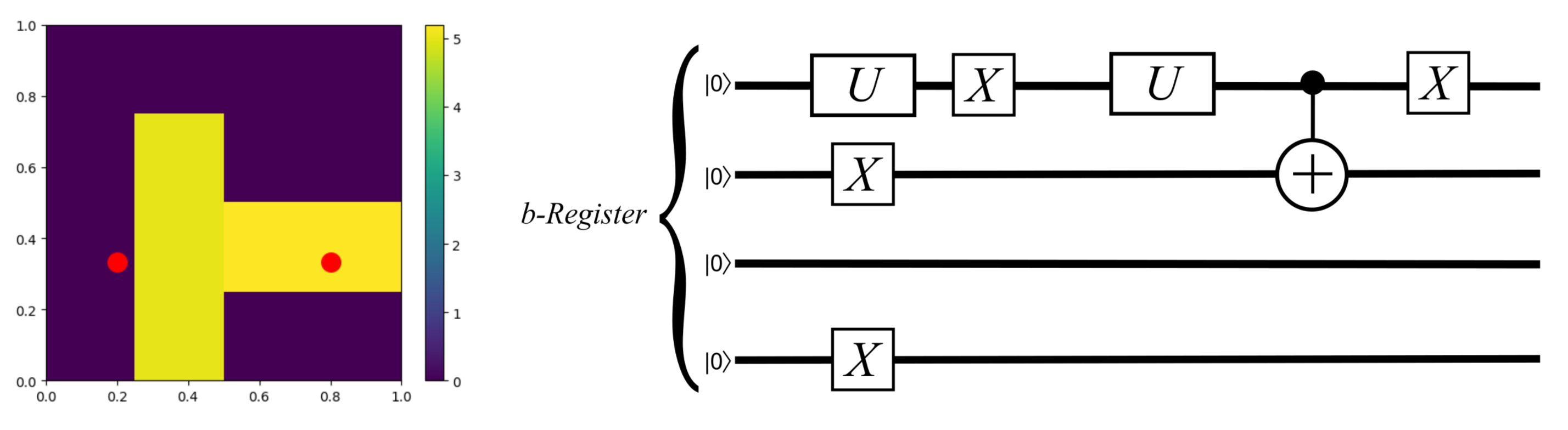}
\caption{On the left, a simplified two-dimensional fracture network model with two fractures intersecting and two random fluid injection/extraction wells at the red dots. The color bar represents permeability on a logarithmic scale. $\mathbf{b}$ encodes boundary conditions for a $4 \times 4$ grid with $2^{n_b}=16$ cells $\Rightarrow{n_b}=4$. Nonzero entries in $\mathbf{b}$ correspond to well sites. On the right, a quantum circuit to prepare the state $|b\rangle=\frac{1}{\sqrt{39665}}(-164|0001\rangle+113|1100\rangle)$ for this fracture flow problem with Neumann boundary conditions (sparse nonzero entries in $\mathbf{b}$).}
\label{fig:sparse_state_prep_ex}
\end{figure}

Consider $\mathbf{b}=(0,-164,0,0,0,0,0,0,0,0,0,0,0,113,0,0)$ where ${n_b}=4$ and there are $W=2$ nonzero coefficients. In this case, we have
\[
|b\rangle=\frac{1}{\sqrt{39665}}(-164|0001\rangle+113|1100\rangle).
\]
By applying the algorithm of Ref. [\citenum{niels_torsten_aeafsqsp}], we can generate this state using the circuit depicted in Figure \ref{fig:sparse_state_prep_ex}.

\begin{figure}[!ht]
\centering
\includegraphics[width=0.5\linewidth]{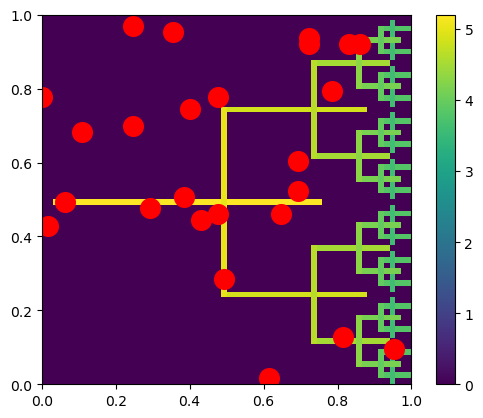}
\caption{An illustration of a two-dimensional fracture network model with fractures that intersect in a pattern of fractal-style recursion to generate a pitchfork fracture network. Random fluid injection/extraction wells are located at the red dots. The color bar represents permeability on a logarithmic scale. $\mathbf{b}$ encodes Neumann boundary conditions (sparse nonzero entries) for a $64 \times 64$ grid with $2^{n_b}=4096$ cells $\Rightarrow{n_b}=12$. Nonzero entries in $\mathbf{b}$ correspond to well sites.}
\label{fig:random_injectors_ex}
\end{figure}

Figure \ref{fig:random_injectors_ex} depicts a two-dimensional fracture network model involving two fractures intersecting in a $\dagger$-configuration, with fractal-style recursion of the $\dagger$-system to generate a more complicated pitchfork fracture network.
The relative permeability of the fractures as compared to that of the underlying rock is a critical parameter in the analysis of fracture systems, and thus, low and high permeability contrast is represented by a gradient scale, where the smallest fractures have the least permeability contrast [\citenum{golden2022preconditioning}].
Fluid injection/extraction wells are located randomly at sites corresponding to red dots.
And, as above, solving this fracture network problem with Neumann boundary conditions requires generating a quantum circuit which prepares the state of a sparse vector $\mathbf{b}$.

We evaluate the performance of the method in Ref. [\citenum{niels_torsten_aeafsqsp}] applied to this fracture problem.
To generate random sparse states comprising ${n_b}$ qubits with $W$ nonzero coefficients, we select $W$ distinct basis states $|x\rangle$, where $x \in\{0,1\}^{n_b}$, from a random, uniform distribution, and we form $\phi$ as the superposition of these selected states.
Coefficient values are chosen from a random normal distribution and subsequently normalized to represent the weights given to random injection/extraction wells in the problem domain.
Note that in a uniform superposition, the coefficients of the $W$ selected basis states are equal, and the size of the state-preparation circuit produced remains independent of the exact coefficient values, as long as they are nonzero.

\begin{figure}[ht]
\centering
\includegraphics[width=0.6\linewidth]{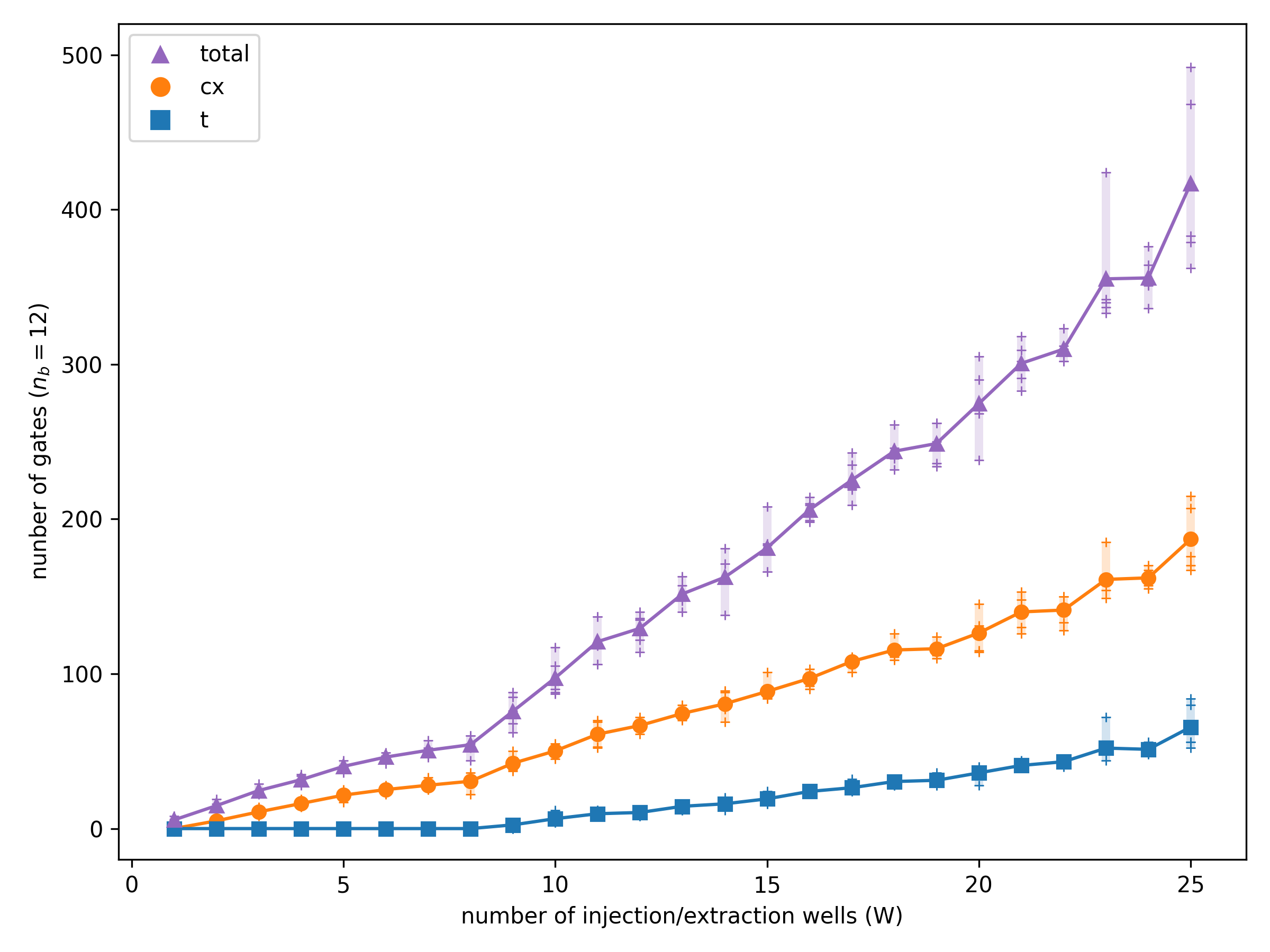}
\caption{Total gate count (total), CNOT gate count (cx) and T gate count (t) of the quantum circuit preparing the state of a sparse vector $\mathbf{b}$.  In our system, the number of qubits is fixed at ${n_b}=12$ while $W$, the number of nonzero entries in $\mathbf{b}$ corresponding to randomly-generated injection/extraction sites, increases from 1 to 25.  For each value of $W$, gate counts from 5 random state samples are shown with $+$ symbols, along with their average and bars extending from smallest to largest value. Theoretically, the state preparation method's total gate count scales as $O(W {n_b})$, further broken down as $O(W n_b)$ CNOT gates and $O(W\log W + n_b)$ single-qubit gates.}
\label{fig:random_injectors}
\end{figure}

In Figure \ref{fig:random_injectors}, we explore how the circuit size scales with an increase in $W$ from 1 to 25 while keeping ${n_b}$ fixed at ${n_b} = 12$.
That is, the number of randomly-generated injection/extraction points increases from left to right along the horizontal axis. 
For each value of $W$, we conduct 5 random state samples.
The average gate count is illustrated in the figure along with a bar indicating the smallest and largest counts. 
The quantum circuits produced of $O(W {n_b})$ size are asymptotically better than those produced by general state preparation methods of size $O(2^{n_b})$ for $W \ll 2^{n_b}$ (\textit{i.e.}, when $\mathbf{b}$ is sparse).

\section*{Efficient Information Extraction}\label{sec:efficient_information_extraction}

\begin{figure}[!b]
\centering
\includegraphics[width=0.7\linewidth]{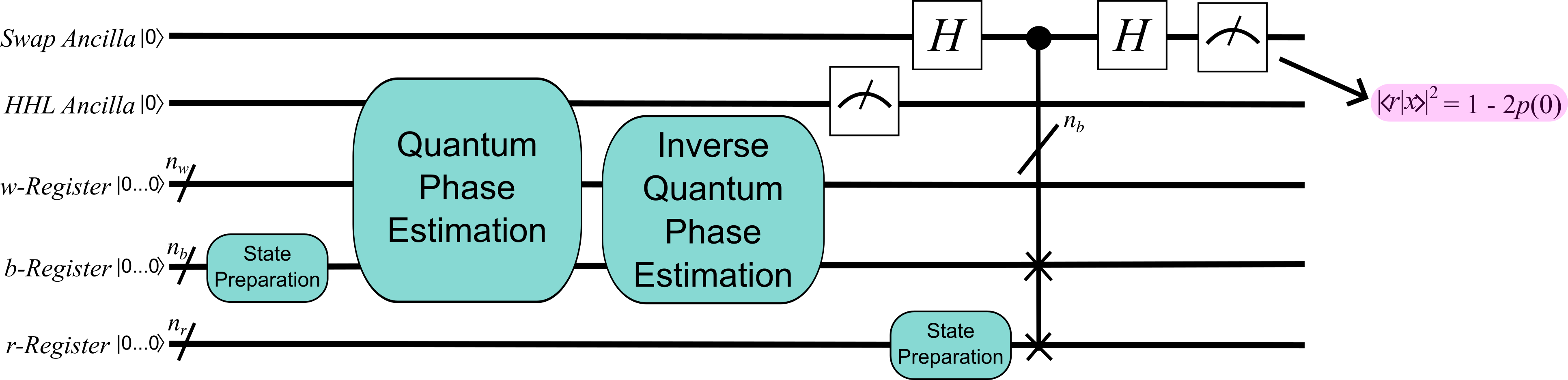}
\caption{A schematic of the circuit structure for extracting average pressure using the swap test. Note that only the magnitude of the average pressure is extracted; sign is not.}
\label{fig:HHL_and_swap}
\end{figure}

\begin{figure}[ht]
\centering
\includegraphics[width=0.7\linewidth]{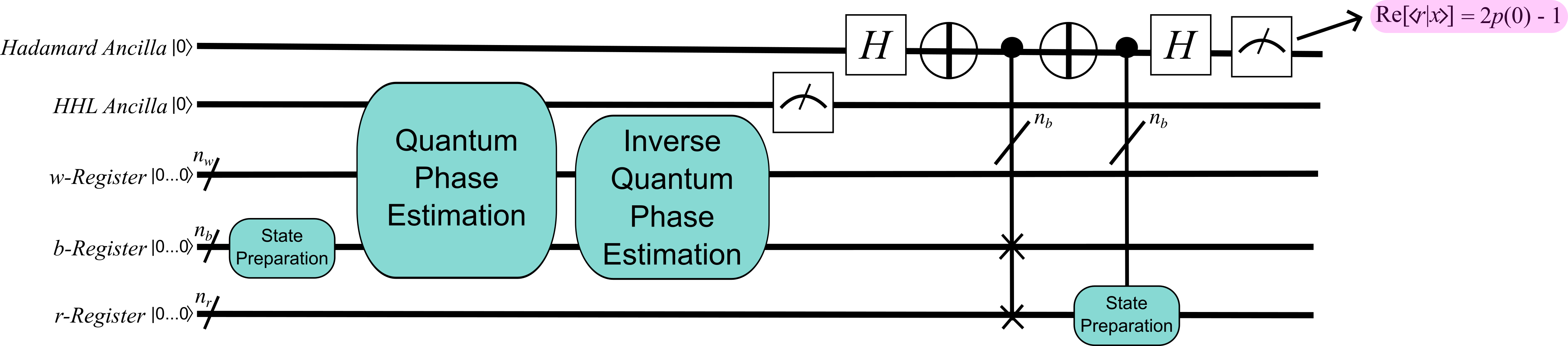}
\caption{A schematic of the circuit structure for extracting average pressure using the Hadamard test. The circuit is more complicated than in Figure \ref{fig:HHL_and_swap}, but it allows for extracting both the sign and magnitude of the average pressure.}
\label{fig:HHL_and_hadamard}
\end{figure}

We now consider a second challenge in applying HHL to geologic fracture systems, namely extracting information upon solution completion.
Because it is unrealistic to obtain all of the pressures from the quantum computer's solution, we must consider other quantities of interest [\citenum{nielsen2010quantum, harrow2009quantum}], and in the realm of geologic fracture networks, one such quantity is the average pressure in a particular region.
Average pressure is a sum of pressures in individual nodes divided by the total number of nodes considered, so we can obtain this using the swap test (or Hadamard test) with a register that prepares an appropriately complementary state to the $\ket{x}$ nodes whose average we seek.
Specifically, consider an $r$-register with at most $n_b$ qubits, where $n_b$ is---as above---the number of qubits in the $b$-register.
If that $r$-register prepares a state such that $\braket{r|x}$ provides the sum of the pressures in a set of nodes, we can obtain $\braket{r|x}$---or at least $|\braket{r|x}|$---via either the Hadamard or swap tests and can then divide by the number of nodes that are in our desired region.
Figures \ref{fig:HHL_and_swap} and \ref{fig:HHL_and_hadamard} illustrate the structure of circuits for extracting information with the swap and Hadamard tests, respectively.
The remainder of this section describes swap test information extraction, Hadamard test information extraction, and a procedure for generating $r$ states to obtain the average pressure for any user-specified region.

First, the swap test approach.
After preparing $\ket{x}$ using HHL, we apply the swap test using an additional ancilla qubit and an $r$-register that contains a state to provide $|\braket{r|x}| = \sum_{i \in \text{user\_elems}} x_i$, where \texttt{user\_elems} is the set of node indices in the region for which average pressure should be computed.
By determining the probability of measuring the swap test ancilla as zero, we can use a classical computer to obtain the unsigned average pressure as follows:
Use $p(0)$ to obtain $|\braket{r|x}|^2 = 1 - 2p(0)$, take the square root of $|\braket{r|x}|^2$, and plug into $\text{avg\_press} = \frac{|\braket{r|x}|}{|\text{user\_elems}|} = \frac{\sum_{i \in \text{user\_elems}} x_i}{|\text{user\_elems}|}$.
The swap test itself (not including the resources required for building the $r$-register, which are analyzed below) requires only one ancilla, two Hadamard gates, and $n_r$ controlled-swap gates, where $n_r$ is the number of qubits required for the $r$-register.
Consequently, if we can build the $r$-register state efficiently, this is an acceptable approach to computing the average pressure in a specified region.\footnote{It is also worth noting that all of these gates are applicable to near-term devices, and the swap test's complexity of $O(1/\epsilon^2)$ for a user-desired error, $\epsilon$, is widely recognized and applied as reasonable for the near-term era, as well [\citenum{omalley22woodbury}]. So, although HHL is not yet realistic for near-term devices, this swap test approach should be, and it would certainly be for fault-tolerant machines.}

Second, the Hadamard test: we first prepare $\ket{x}$ using HHL, and then use an additional ancilla qubit and an $r$-register to prepare $|\braket{r|x}| = \sum_{i \in \text{user\_elems}} x_i$, where \texttt{user\_elems} is as defined above.
Not counting the resources required for building the $r$-register, this requires two Hadamard gates, two Pauli-$\mathbf{X}$ gates, and at most $n_r$ controlled-swap gates.
It is worth clarifying how the controlled gates work: the controlled-swap gates transfer the components of $\ket{x}$ that will be used in the inner product computation to the $r$-register, and these controlled-swaps thus replace the controlled-$U_\phi$ gate in Subfigure B of Figure \ref{fig:swap_and_hadamard_test}.
The controlled-state preparation gate then prepares the $r$-register, as described below.
So, to obtain the average pressure, we compute the probability of measuring the Hadamard test ancilla as zero, and use it to classically compute $\text{Re}[\braket{r|x}] = 2p(0) - 1$.
We then divide by the number of desired nodes to find $\text{avg\_press} = \frac{\text{Re}[\braket{r|x}]}{|\text{user\_elems}|} = \frac{\sum_{i \in \text{user\_elems}} x_i}{|\text{user\_elems}|}$.

\subsection*{Building the $r$ Register}
Using either the swap or Hadamard tests requires an $r$-register that properly extracts sums of equally-weighted node values.
For the remainder of this section, we will illustrate the $r$-register-building process for the swap test, but using the Hadamard test would require only two modifications.
First, instead of computing $|\braket{r|x}|^2$ using the probability that the ancilla is zero, we would compute $\text{Re}[\braket{r|x}]$, which does not require taking a square root.
Second, while the $r$-register preparation for the swap test does not require controlled gates, each of the gates used to build the $r$-register for the Hadamard test would need to be controlled.

To build the $r$-register, we must determine both the number of qubits, $n_r$, and what gates need to be applied to these qubits such that $|\braket{r|x}|^2$ is a sum of the form $|\braket{r|x}| = \xi\sum_{i\in\text{user\_elems}} x_i$.
If the quantum computer can provide a $|\braket{r|x}|^2$ with that form, then, as long as we know $\xi$, we can use the few classical computations described above to compute the average pressure in the set of nodes specified by \texttt{user\_elems}.
It is worth noting that our approach does not treat the initial state of the qubits in the $r$-register as a variable to be determined; the initial state of each $r$-register qubit is always the fiduciary state, $\ket{0}$.

\begin{figure}[t]
\centering
\includegraphics[height=0.2\linewidth]{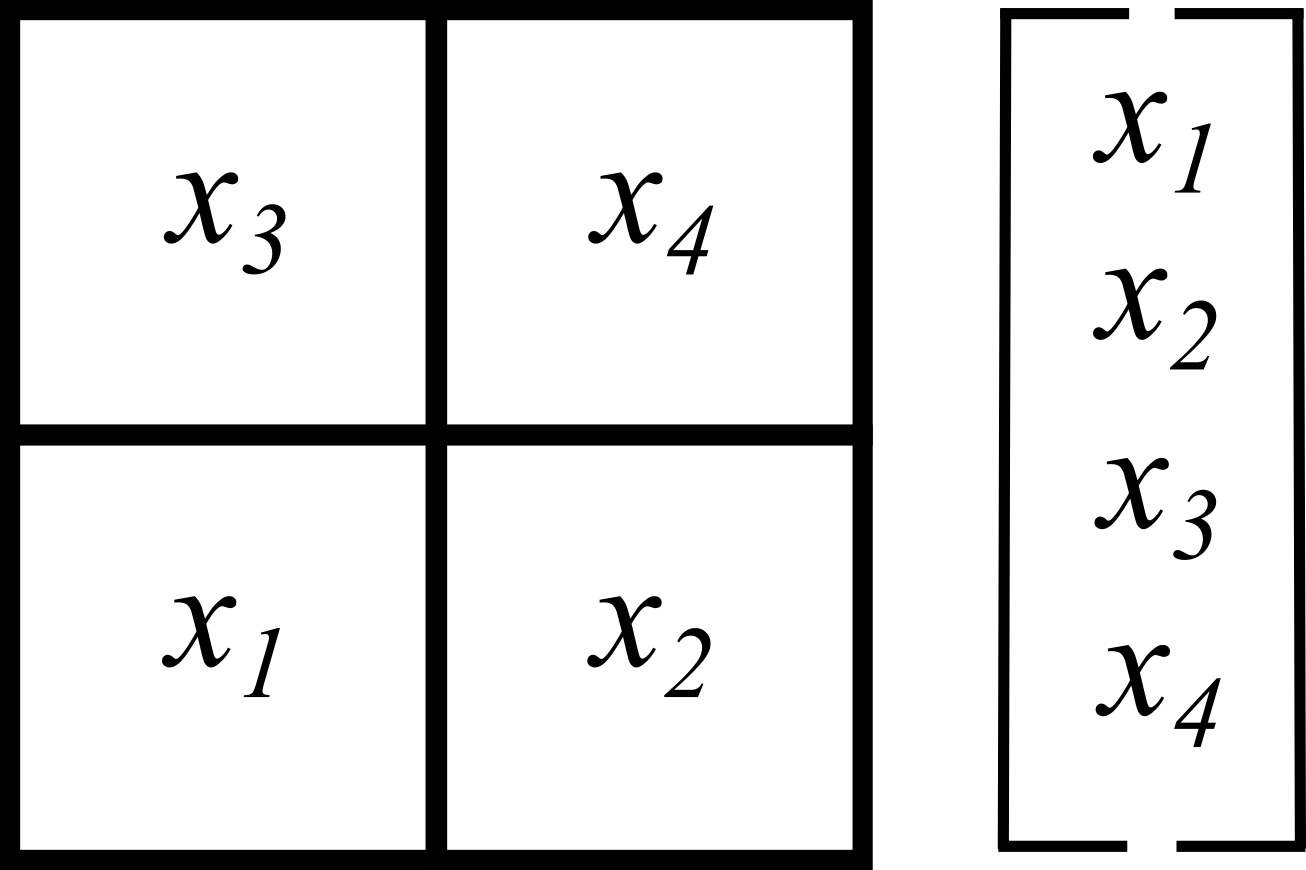}
\caption{A schematic $2 \times 2$ grid illustrating a discretized region with four nodes, some combination of which holds the average pressure we seek to compute.}
\label{fig:two_by_two_grid}
\end{figure}

While there is a general procedure for building $r$-registers that obtain the average pressure in an arbitrarily-specified set of discretized nodes, special cases offer simpler methods that have fewer steps and that aid understanding of the general process.
Therefore, we first consider the special cases of an entire row, an entire column, and an entire diagonal of discretized nodes using the example region of Figure \ref{fig:two_by_two_grid}, which is the simplest possible discretized region that satisfies the constraints of being square and having a number of nodes that is a power of two.
It corresponds to a $b$-register of $n_b = 2$ qubits, and after applying the HHL algorithm, the final states of those qubits are $x_1\ket{0} + x_2\ket{1}$ and $x_3\ket{0} + x_4\ket{1}$, where $\mathbf{x} = \gamma \begin{bmatrix}
x_1 \\ x_2 \\ x_3 \\ x_4 \end{bmatrix}$ with a normalization constant $\gamma$.
Because two nodes' worth of information are stored in each qubit, we will term a `pair' of nodes to be any whose information is stored on the same qubit.
Thus, using the node-indexing scheme of Figure \ref{fig:two_by_two_grid}, we have pairs $x_1$, $x_2$ and $x_3$, $x_4$.
When both nodes in a pair are included in the desired average pressure region, we will term that group of nodes `paired nodes.'
When a node is included in the desired region without its paired node, we will term that node a single node.
Nodes with even indices will be termed `even-indexed,' and vice versa for nodes with odd indices.

The discretized region of Figure \ref{fig:two_by_two_grid} has ten possible average pressures: the average pressure from a single node (either $x_1$, $x_2$, $x_3$, or $x_4$), from row $x_1$ + $x_2$, from row $x_3$ + $x_4$, from column $x_1$ + $x_3$, from column $x_2$ + $x_4$, from diagonal $x_1$ + $x_4$, and from diagonal $x_2$ + $x_3$.
First, consider the case of a single node; the `average' pressure of a single node is, of course, simply the pressure in that node.
Consequently, an $r$-register-generation procedure for obtaining the value in a single node has little utility.
Not only is the primary goal of extracting average pressure to avoid the costly procedure of sequentially extracting single elements from the solution, but it would also likely be simpler to just measure qubits of interest if the goal were to extract a few nodes' worth of individual pressures.
Therefore, we will not further discuss the single node case.

Second, consider the rows; both cases require that the $r$-register prepare a state such that that $|\braket{r|x}| = r_ix_i + r_{i+1}x_{i+1}$ can be simplified to $\xi(x_i + x_{i+1})$, for $\xi=r_i=r_{i+1}$ and row nodes $x_i$ and $x_{i+1}$.
Note that rows are thus comprised of only paired nodes, and applying a Hadamard gate to each qubit in the $r$-register lends such an equal $\xi$ coefficient to both nodes in a pair. 
So, to build the $r$-register for an entire row, add a qubit for each pair of nodes in the row, and apply a Hadamard gate to each qubit.
This allows for computing the pressure in, say, the row with $x_1$ and $x_2$ as follows:
\begin{equation}
\begin{split}
     |\braket{r|x}|^2 = \left(\frac{1}{\sqrt{2}}(x_1 + x_2)\right)^2 \\
     \Rightarrow \text{avg\_pressure} = \sqrt{2}\frac{|\braket{r|x}|}{2}.
\end{split}
\end{equation}
Consequently, the $r$-register preparation circuit for a row requires one qubit for every pair of nodes in the row and one Hadamard gate for every pair of nodes in the row.

Second, consider the case of a column, which has two subcases.
The first subcase has columns with only odd-index nodes, and the second has columns with only even-index nodes; there will never be paired nodes in a column.
For the even-index subcase, the $r$-register must use Pauli-$\mathbf{X}$ gates to flip the state of the $r$-register qubits from the initial state of $\ket{0}$ to a state of $\ket{1}$, such that a coefficient of $0$ will be multiplied with each of the odd-indexed $x$ values in a given qubit pair, and a coefficient of $1$ will be multiplied with each of the even-indexed $x$ values in the same.
In the odd-index subcase, although we need one $r$-register qubit per desired node in the average pressure computation, that qubit need not have any gates on it.
This is because every qubit in the $r$-register is initialized to the state $\ket{0}$, so a coefficient of $1$ will already be multiplied with each of the odd-indexed $x$ values whose average we seek.
Consequently, in the column case, we need one qubit for every node in the desired region and one Pauli-$\mathbf{X}$ gate for every even-indexed node in the same.
Then, the pressure in column $x_1$ and $x_3$ could be computed as 
\begin{equation}
\begin{split}
     |\braket{r|x}|^2 = \left((1)x_1 + (1)x_3\right)^2 \\
     \Rightarrow \text{avg\_pressure} = \frac{|\braket{r|x}|}{2}.
\end{split}
\end{equation}

\begin{figure}[ht]
\centering
\includegraphics[width=\linewidth]{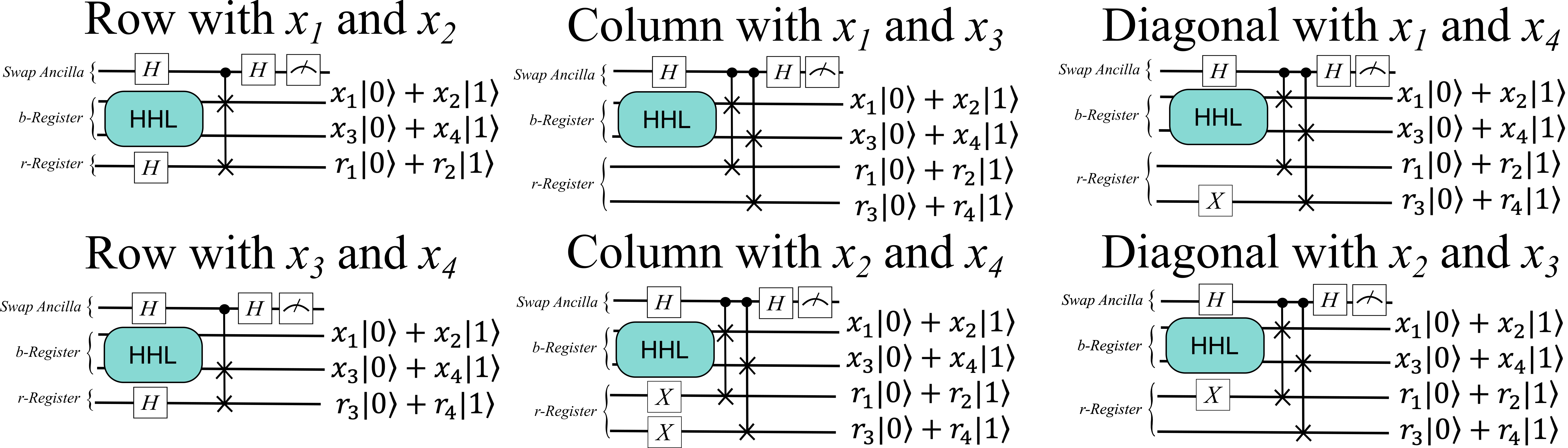}
\caption{Schematic of the HHL and swap test circuits for extracting the average pressure from a full row, column, and diagonal of a discretized two-by-two region.}
\label{fig:simple_r_register_generation}
\end{figure}

Third is the full diagonal case, which is functionally-equivalent to that of the column case, because a discretized diagonal will never involve paired nodes.
Consequently, the $r$-register will again be comprised of one qubit per desired node in the average-pressure region, with one Pauli-$\mathbf{X}$ gate on each $r$-register qubit being swapped with an even node.
Figure \ref{fig:simple_r_register_generation} summarizes the $r$-register generation procedures thus far.

We can extend these patterns to the more general case of arbitrarily-selected nodes in which to obtain the average pressure.
We use the same patterns of applying a Hadamard gate on $r$-register qubits with paired node indices and a Pauli-$\mathbf{X}$ gate on $r$-register qubits with single, even-indexed nodes in the desired region.
But now, obtaining average pressure in some sets of nodes will require more than one swap or Hadamard test.
Specifically, in regions that contain both paired and single nodes, we will need two swap tests or two Hadamard tests: one test for paired nodes will result in a value with a coefficient of $\frac{1}{\sqrt{2}}$, while the other test for single nodes will produce a value with a coefficient of one.
We can then compute average pressure by multiplying the result of the test for paired nodes by $\sqrt{2}$ to remove the coefficient, adding the resulting sum with the sum that results from the second test for single nodes, and dividing by the total number of nodes in the desired region.

\begin{figure}[!b]
\centering
\includegraphics[width=\linewidth]{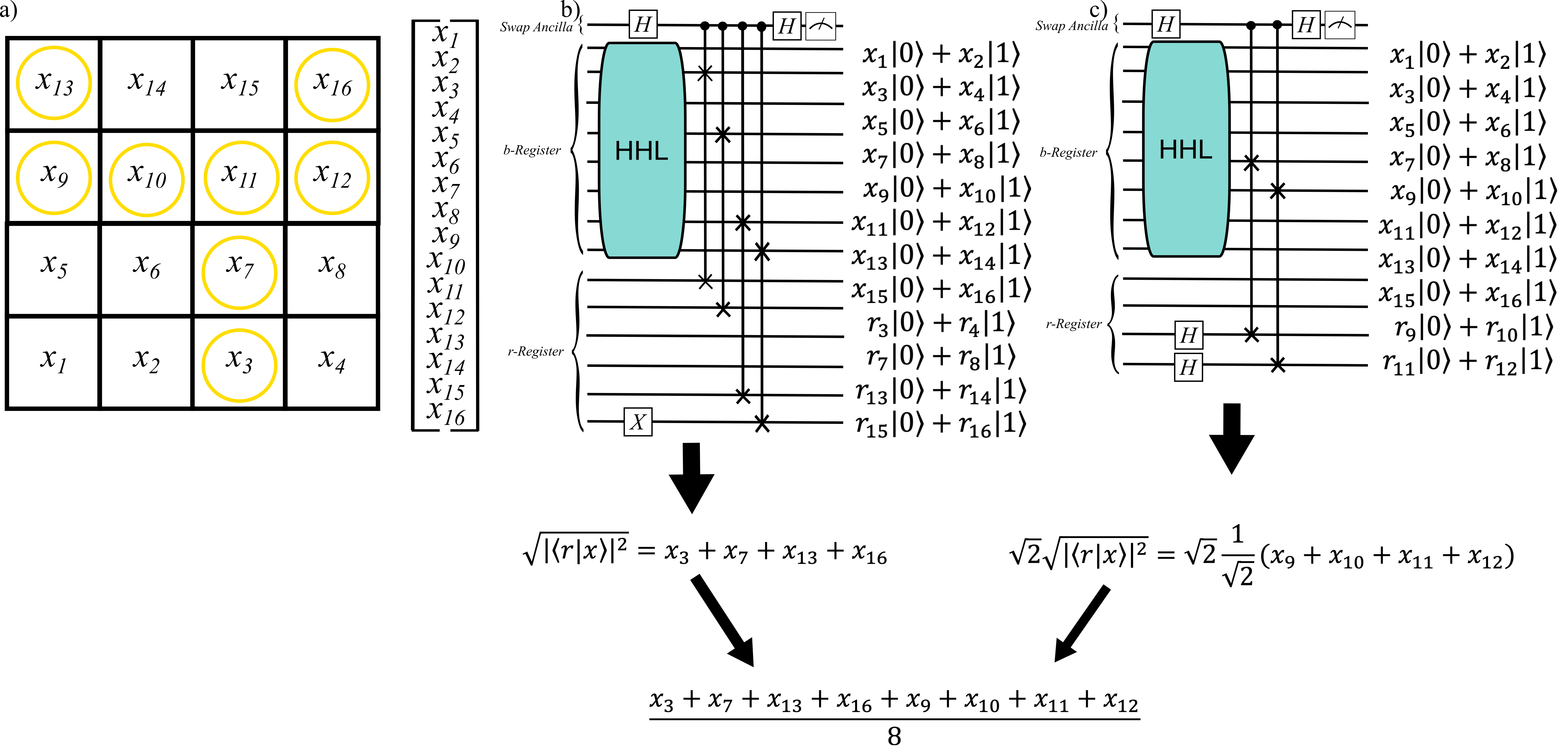}
\caption{Subfigure A is a $4\times 4$ discretized region with a pitchfork fracture marked by yellow circles. Subfigure B illustrates the two circuits needed to compute the average pressure in a series of arbitrarily-selected nodes, by building $r$-registers as described above.}
\label{fig:four_by_four_region}
\end{figure}

To clarify, consider a slightly more complicated example where we have regions that are not automatically an entire row, column, or diagonal, as they were in the case of Figure \ref{fig:two_by_two_grid}.
Specifically, consider the $4\times 4$ discretized region in Subfigure A of Figure \ref{fig:four_by_four_region}, and suppose that the nodes circled in yellow represent nodes through which a two-pronged pitchfork fracture runs.
Obtaining the average pressure in this fracture requires two circuits as illustrated in Subfigure B of Figure \ref{fig:four_by_four_region}.
The first computes the sum of pressures in nodes that are single, while the second computes the sum of pressures in nodes that are pairs.
Again, the difference between the circuit outputs is that paired nodes require multiplying by a known factor of $\sqrt{2}$ to remove the $\frac{1}{\sqrt{2}}$ term resulting from the Hadamard gates in the $r$-register.
After obtaining the sums, we can classically compute the average by dividing by the known number of nodes in the desired region.

Because building $r$-registers for arbitrarily-specified regions uses the same procedures as for building the registers for entire rows, columns, or diagonals, the general $r$-register complexity also aligns with the complexities of those situations.
Specifically, the circuit for determining the average pressure of single nodes requires an $r$-register with one qubit per desired node and one Pauli-$\mathbf{X}$ gate per even-indexed node, while the circuit for determining the average pressure of paired nodes requires an $r$-register with one qubit per pair of nodes and one Hadamard gate per pair of nodes.

\subsection*{Overall Complexities: Qubit Count, Gate Count, and Additive Error for Average Pressure Extraction}
The complexity in terms of qubit and gate count for the HHL algorithm is well-understood [\citenum{harrow2009quantum,chakrabarti18,duan2020survey}], and is taken to be acceptable, if not for near-term machines, then for fault-tolerant or error-mitigated machines of the future [\citenum{kim2023evidence}].
Therefore, to assess whether our information extraction approach is appropriately efficient, we assess its worst-case complexity in terms of qubit and gate counts that are already required for the HHL portion of the circuit.
We also comment on the relationship between the additive errors introduced during HHL and the information extraction portions of the computation.

First, qubit count:
both the swap and Hadamard tests require only $n_r + 1$ additional qubits, so we need to determine how $n_r$ scales with the size of the circuit required for HHL.
Whether we have a region with all single nodes or all paired nodes, the maximum number of qubits in the $r$-register is $n_b$. 
Furthermore, the total number of $r$-register qubits cannot be greater than $n_b$ even across two circuits for single and paired nodes, because if it were, at least one node would be double-counted in the resulting average.
Therefore, $n_r$ is $O(n_b)$, and the overall number of additional qubits for the swap/Hadamard test is $O(n_b) + 1 = O(n_b)$.
As $n_b$ qubits is considered reasonable in the HHL algorithm, an additional $O(n_b)$ is a reasonable worst-case-scenario for information extraction.

Second, gate count; both the swap and Hadamard tests add a few additional gates (two for the swap test and four for the Hadamard test) in addition to the controlled-swap gates and the gates required to construct the $r$-register.
Because there can be at most $n_b$ qubits in each $r$-register, there can be at most $n_b$ controlled-swap gates, meaning the worst-case-scenario for controlled-swap gates is $O(n_b)$.
Additionally, we found that the maximum number of gates required in the $r$-register is also $O(n_b)$ because at most one additional gate (either a Hadamard or a Pauli-$\mathbf{X}$) is required per qubit in the $r$-register.
Consequently, the overall number of gates is $O(n_b)$ (either $2 + O(n_b) + O(n_b)$ or $4 + O(n_b) + O(n_b)$), which is---by the same reasoning as above---considered efficient for purposes of HHL and therefore also for purposes of information extraction.

We note that a more interesting complexity issue is raised by the two circuits required to obtain average pressure in a region with both single and paired nodes.
This requires twice as many executions of the entire HHL procedure, which---particularly in the contemporary quantum hardware ecosystem---may add up to substantial runtime expense.
Therefore, it is worth noting that we can avoid using two circuits to compute the average pressure in an arbitrary region, if we are willing to select only single nodes to represent our desired region.
For example, consider again the fracture in Subfigure A of Figure \ref{fig:four_by_four_region}; if we removed nodes $10$ and $12$---or nodes $9$ and $11$ or $10$ and $11$ or $9$ and $12$---then we could use a single circuit with the structure of Figure \ref{fig:four_by_four_region}'s Subfigure B to compute the average pressure in a region comprised of only single nodes.
While this does not change the overall possible complexity of that circuit, it does reduce circuit executions to half of what would otherwise be required, at an `accuracy cost' of removing just a few nodes from our desired average pressure region.

Third and finally, error propagation.
The HHL algorithm prepares a solution $\ket{x}$ with an additive error of $\epsilon_{HHL}$, where $\epsilon_{HHL}$ is a user-defined parameter that will affect the QPE and inverse QPE subroutines [\citenum{harrow2009quantum, chakrabarti18}].
Therefore, upon completion of HHL, the $b$-register will be in a state with every element of $\ket{x}$ plus at most $\epsilon_{HHL}$.\footnote{For simplicity, we assume that every element of the solution vector has the same error, which could be interpreted as the maximum of individual element errors, and which lends a state $\ket{x + \epsilon_{HHL}}$.}
The swap and Hadamard tests both introduce a second additive error, $\epsilon_{extract}$, such that the inner product measured has a value of at most $\epsilon_{extract}$ added to it.

Consider the structure of our $r$-registers, which weight the elements whose average we seek \textit{equally}.
This means that each node in the average will introduce an error of $1*\epsilon_{HHL}$, for a total of $n*\epsilon_{HHL}$, where $n$ is the number of desired nodes in the average computation.
Then, an error of $\epsilon_{extract}$ will be added to the entire inner product---including the $n*\epsilon_{HHL}$ term---lending an overall error of $n*\epsilon_{HHL}$ + $\epsilon_{extract}$.

Because both additive errors are specified by the user and relate to the number of circuit executions required, the user can determine the amount of error in the average pressure by choosing a number of circuit executions they want to perform and using the complexities of HHL and the swap test ($O(\log{(N)}\kappa s^2/\epsilon_{HHL})$ and $O(1/\epsilon_{extract}^2)$, respectively) to `back out' approximate values of $\epsilon_{HHL}$ and $\epsilon_{extract}$.
Then, the overall additive error in the average pressure solution is $n*\epsilon_{HHL}$ + $\epsilon_{extract}$, which is $O(\epsilon)$, for $\epsilon=\max{\{\epsilon_{HHL},\epsilon_{extract}\}}$.
Assuming that $\epsilon_{HHL}$ and $\epsilon_{extract}$ are chosen to have similar magnitudes, our approach to efficient information extraction thus has an overall error that is the same order as the additive error of HHL itself, meaning our information extraction procedure is consistent with errors that are widely viewed as acceptable.

\section*{Conclusion}
This work introduces efficient methods for both preparing the $b$-register for and extracting useful information from a quantum computer that solves a linear system representing a geologic fracture network problem.
Both approaches have reasonable complexities, as they require $O(n_b)$ additional qubits and either $O(W n_b)$ or $O(n_b)$ additional gates, where $W$ is the number of injection/extraction wells.
Furthermore, both approaches utilize only straightforward gate types that are applicable to near-term hardware  [\citenum{omalley22woodbury}], making them viable on improved hardware of the future.
Consequently, this work---combined with the previous results in Refs. [\citenum{henderson2023quantum}] and [\citenum{golden2022preconditioning}]---provides answers to three of the four ``fine print'' considerations for using quantum algorithms to solve geologic fracture network problems.
To be clear, these works do not close inquiry into the issue of solving geologic fracture network problems with quantum algorithms; for example, there are likely other solutions to state preparation and information extraction considerations, some of which may lend themselves more readily to quantum linear solvers that are more applicable for today's noisy, near-term devices.
Rather, this work and those it complements provide a foundational answer to the question of using quantum computing in the geologic fracture realm: it can be done and, especially with the rapid ascent of ever-improving quantum hardware [\citenum{choi2022ibm}], has the promise to revolutionize the modelling of geologic fracture networks.

\bibliography{references}

\section*{Acknowledgements}

JMH and DO gratefully acknowledge support from the Department of Energy, Office of Science, Office of Basic Energy Sciences, Geoscience Research program under Award Number (LANLE3W1).
JMH, JKG, and DO gratefully acknowledge support from Los Alamos National Laboratory's Laboratory Directed Research \& Development project 20220077ER.

\section*{Author contributions statement}

JKG and DO designed the project.
JK, AGP, and DO designed the approach for efficient state preparation, and JK implemented the associated codes.
JMH, JKG, and DO designed the approach for efficient information extraction.
JMH and JK prepared the first draft of the manuscript.
All authors reviewed and revised the manuscript. 

\section*{Additional information}

\subsection*{Data Availability Statement}
The codes---which generate the considered data---are available at https://github.com/OrchardLANL/DPFEHM.jl/tree/master\\/examples/fracture\_networks\_for\_qc.

\subsection*{Competing Interests}
The authors declare no competing interests.

\end{document}